# Addressing Weak Authentication like RFID, NFC in Electric Vehicle (EVs) and Electric Vehicle Charging Systems (EVCs) using AI-powered Adaptive Authentication

**Onyinye Okoye**

**University of Denver**

# Abstract

The rapid expansion of the Electric Vehicles (EVs) and Electric Vehicle Charging Systems (EVCs) has introduced new cybersecurity challenges, specifically in authentication protocols that protect vehicles, users, and energy infrastructure. Although widely adopted for convenience, traditional authentication mechanisms like Radio Frequency Identification (RFID) and Near Field Communication (NFC) rely on static identifiers and weak encryption, making them highly vulnerable to attack vectors such as cloning, relay attacks, and signal interception. This study explores an AI-powered adaptive authentication framework designed to overcome these shortcomings by integrating machine learning, anomaly detection, behavioral analytics, and contextual risk assessment. Grounded in the principles of Zero Trust Architecture, the proposed framework emphasizes continuous verification, least privilege access, and secure communication. Through a comprehensive literature review, this research evaluates current vulnerabilities and highlights AI-driven solutions to provide a scalable, resilient, and proactive defense. Ultimately, the research findings conclude that adopting AI-powered adaptive authentication is a strategic imperative for securing the future of electric mobility and strengthening digital trust across the ecosystem.

*Keywords:* weak authentication, RFID, NFC, ML, AI-powered adaptive authentication, relay attacks, cloning, eavesdropping, MITM attacks, Zero Trust Architecture

# Table of Contents

# Illustrations

**Figures**

**Tables**

**Chapter 1: Background**

Problem Statement

The rapid adoption of Electric Vehicles (EVs) has driven significant growth in Electric Vehicle Charging Systems (EVCs) and deepened integration with the power grid. EVs are equipped with advanced connectivity features and multiple security access points, enabling seamless interaction with charging stations, users, and the energy infrastructure. As the electric mobility ecosystem continues to evolve, standardizing robust security protocols is crucial to ensure secure adoption, protect digital infrastructure, and maintain the sustainability of the power grid.

A key component of these protocols is authentication, critical for securing vehicle access, protecting user data, and preventing unauthorized usage. Currently, technologies such as Radio-Frequency Identification (RFID) and Near-Field Communication (NFC) are the most widely implemented authentication mechanisms in EVs and EVCs. These contactless technologies are favored for their convenience and are governed by international standards such as ISO 14443 and ISO 15693. However, these standards rely on plaintext transmission and weak cryptographic methods, making them vulnerable to security threats such as cloning, relay attacks, and signal interception (Garfinkel, Juels, & Pappy 2005).

Despite their widespread use, RFID and NFC present considerable security risks. Attack vectors include man-in-the middle (MITM) attacks, data modification, denial-of-service (DoS), identify spoofing, and encryption compromise. These vulnerabilities can lead to serious consequences, including financial loss, data breaches, energy theft, and infrastructure disruption (Shi et al. 2025). To address these issues, emerging solutions incorporate enhanced encryption, cryptographic response protocols, multifactor authentication, and increasingly, AI-powered adaptative authentication systems. This study evaluates effectiveness of AI-powered adaptative authentication protocols in securing EVs and EVCs. It explores how artificial intelligence, through pattern recognition, anomaly detection, and predictive analytics can proactively identify and mitigate evolving cyber threats. Ultimately, this paper argues that AI-driven authentication provides a more secure and scalable alternative to traditional RFID and NFC methods, offering a robust framework for securing digital identities within the electric mobility ecosystem.

<div align="center">Importance of Topic</div>

The selection of this capstone project is driven by a strong interest in the intersection of EVs, EVCs, and cybersecurity. As EVs and EVCs increasingly function as "computers on wheels", it is critical to explore both the rapid evolution of these technologies and the growing sophistication of cyber threats targeting their interconnected systems, particularly the vulnerabilities in EVs and EVC authentication methods.

The widespread adoption of Artificial Intelligence (AI) has further escalated the cybersecurity landscape by equipping threat actors with powerful tools to launch advanced cyberattacks and compromise data privacy. This heightened risk underscores the urgency for innovative, AI-driven security solutions. The capstone will explore the practical application of AI-powered adaptive authentication to identify key security threats and develop a framework for more secure and responsive access control mechanisms. Such work aims to help the automotive industry to stay ahead of threat actors and protect both users and critical infrastructure.

Undertaking this project offers an opportunity to demonstrate applied knowledge in cybersecurity and artificial intelligence, while also enhancing skills such as research, critical thinking, data analysis, and creative problem solving. Ultimately, this research will deepen my understanding of emerging technologies and security challenges in the automotive sector and contribute to the advancement of robust cybersecurity policies and authentication systems for EV and EVC infrastructure.

<div align="center">Thesis Statement</div>

This research proposes an AI-powered adaptive authentication framework for EVs and EVC systems. The framework will leverage machine learning and real-time threat intelligence to identify and address authentication vulnerabilities, such as RFID and NFC technologies, in EVs and EVCs. The framework aims to enhance security, protect user privacy, and standardize secure access across the EV ecosystem.

<div align="center">Definition of Terms</div>

This section defines the key terms referenced in the study to ensure clarity and a shared understanding between the author and the audience.

*AI-powered adaptive authentication*

This advanced mechanism leverages artificial intelligence to dynamically adjust identity verification requirements in real time. Rather than relying on static credentials, it continuously analyzes contextual factors such as user behavior, geographic location, time of access, and device health to generate a risk score. Based on this score, the mechanism adapts its response, granting seamless access for low-risk scenarios, requiring additional verification for moderate risk, or blocking access entirely in high-risk situations (Durgaraju 2024).

*Machine learning*

Machine learning is a subset of artificial intelligence (AI) in which algorithms are trained on large datasets to identify patterns, make predictions, and improve performance over time without being explicitly programmed for each task. In context of this paper, machine learning models can be trained on datasets from EVs and EVCs to learn the characteristics of legitimate authentication events. By establishing these patterns of normal behavior, the system can detect anomalies that may indicate cyberattacks, such as relay attack or identity spoofing (Chauhan et al,2024).

*Zero Trust Architecture (ZTA)*

Zero Trust is a modern security framework built on the principle of 'never trust, always verify.' It recognizes that threats may originate both inside and outside the network perimeter, thereby eliminating the concept of implicit trust. Under a Zero Trust Architecture, every request for access to resources must be strictly authenticated and authorized, regardless of its source. This model enforces least privilege access and continuous verification, standing in contrast to traditional security approaches that grant trust once a user or device is inside the network. In this paper, the proposed AI-powered adaptive authentication framework is positioned as a key enabler for implementing Zero Trust in the EV ecosystem (NIST. 2020).

Chapter Summary

This chapter examines the increasing need for robust authentication in EVs and EVCs as they become more tightly integrated with the power grid. Traditional static models such as RFID and NFC, while convenient, present vulnerabilities including cloning, relay attacks, and signal interception. To mitigate these risks, the chapter proposes AI-powered adaptive authentication as a more secure alternative. Leveraging machine learning for continuous anomaly detection and predictive analytics, this adaptive model can proactively identify and respond to evolving cyber threats. Grounded in Zero Trust principles, the framework enforces continuous verification to safeguard user data, vehicle access, and grid security.

## Chapter 2: Approach

Research Approach

This study employs a qualitative research design to explore an AI-powered adaptive authentication framework for Electric Vehicles (EVs) and Electric Vehicle Charging (EVC) systems. This approach focuses on developing innovative, real-world solutions while contributing to academic understanding of EV security. The research is guided by the following key questions: (1) What are the major security vulnerabilities associated with current RFID and NFC authentication methods in EV infrastructure? (2) How can AI-based authentication systems address these vulnerabilities more effectively?

The project will begin with a comprehensive literature review of existing authentication technologies used in EVs and EVCs, focusing on their technical specifications, advantages, and limitations. Peer-reviewed journals, technical white papers, and cybersecurity incident reports will serve as the primary sources.

The research will be conducted over a ten-week period and will culminate in a set of actionable recommendations to strengthen security protocols. These recommendations are intended to support the secure adoption of electric mobility technologies and ensure the long-term sustainability and resilience of the automotive and energy industries.

Research Model

A comprehensive investigation will be carried out to analyze the vulnerabilities associated with various authentication protocols and to compare the features of RFID, NFC and AI-powered adaptative authentication protocols. This study examines multiple dimensions of authentication systems across security and reliability dimensions, to ensure that the adaptative authentication framework effectively balances the security framework with dependable performance. The evaluation model will be guided by the principles outlined in the ISO/IEC 42001 international standard for AI management systems, as well as other industry compliance regulations, to ensure adherence to best practices and regulatory requirements.

Data Collection and Analysis

The study leverages qualitative data collection and analysis techniques to inform the design and development of the AI-powered authentication framework, with an emphasis on enhancing user and system security. The data collection strategy of finding relevant literature began with defining the literature search strategy. A well-constructed search strategy ensured searching for evidence included in the literature review (Bramer et al. 2018). The strategy involved researching, locating, and reviewing conceptual, theoretical, and empirical peer-reviewed literature. A detailed review of the literature focused specifically on the weak authentication like RFID and NFC-based systems, cybersecurity attacks like replay attacks, Man-in-the-Middle (MITM) attacks, RFID card cloning, and eavesdropping, and AI-powered adaptive authentication.

The Google Scholar database was the primary search engine source used in performing extensive search into locating relevant seminal, contextual, and recently published literature. Academic databases used include IEEE, SAGE, and ScienceDirect. Other electronic databases, such as governmental websites, were also used to locate relevant articles. Peer-reviewed and scholarly sources identified from the search results obtained through the search terms and combination of words included 'weak authentication,' 'RFID,' 'NFC,' 'machine learning,' 'artificial intelligence,' 'AI-powered authentication,' 'relay attacks,' 'RFID card cloning,' 'eavesdropping,' 'MITM attacks,' and 'Zero Trust Architecture.' Only relevant sources were selected from the search results.

Qualitative data gathered from various sources including peer-reviewed papers, industry datasets, and interviews are reviewed to uncover valuable insights into EVs and EVC systems and behavioral metrics. By combining findings from academic research with real-world perspectives, this approach offers a comprehensive understanding of how AI-powered adaptive authentication can address weaknesses in RFID and NFC authentication and enhance the security of EVs and EVCs.

Result Design

The research design integrates qualitative findings into the development of an adaptive authentication framework that utilizes machine learning techniques. Insights gathered through data collection and analysis will guide the design of the framework, which aims to enhance EV and EVC security by dynamically evaluating user behavior, identifying risk levels, detecting behavioral anomalies, and adjusting authentication requirements in response to emerging threats. This design emphasizes an iterative, data-informed process in which behavioral patterns and risk indicators are continuously assessed to ensure the authentication system remains both secure and response to evolving security challenges within the electric mobility ecosystem.

<div align="center">Chapter Summary</div>

Chapter 2 presented the qualitative research approach to investigate an AI-powered adaptive authentication framework for EVs and EVC systems. A comprehensive review of existing literature was conducted to ground the study in current academic and industry practices, allowing the investigator to identify patterns, gaps, and emerging challenges in authentication security.

Through this approach, the study addressed two guiding research questions: (1) What are the major security vulnerabilities associated with current RFID and NFC authentication methods in EV infrastructure? and (2) How can AI-based authentication systems address these vulnerabilities more effectively? By synthesizing prior findings and critically analyzing technological trends, the chapter provided a foundation for understanding both the limitations of traditional methods and the potential of AI-driven solutions.

**Chapter 3: Literature review**

Introduction

The primary purpose of this literature review section is to introduce the reader to authentication technologies in Electric Vehicles (EVs) and Electric Vehicle Charging Systems (EVCs), known vulnerabilities and attack vectors, and AI-powered adaptive authentication. The literature review is based on the presented research questions: What are the major security vulnerabilities associated with current Radio-Frequency Identification (RFID) and Near Field Communication (NFC) authentication methods in EV infrastructure? How can AI-powered authentication systems address these vulnerabilities more effectively? The literature review is organized into four sections. The first section traces the evolution of authentication technologies within the EV ecosystems. The second section analyzes vulnerabilities and attack vectors that exploit traditional authentication methods like RFID and NFC. The third section explores research on AI-powered adaptive authentication as a potential solution to these challenges. Finally, the fourth section examines Zero Trust Architecture as a foundational framework to support Adaptive Authentication. The comprehensive literature discussions within Chapter 3 systematically connect the existing bodies of literature with the qualitative research design and identify gaps in research the study findings could fill.

Evolution of Authentication Technologies in the EV Ecosystem

The global shift toward sustainable energy has catalyzed a transformation in the automotive industry, driven by the rapid adoption of EVs. This evolution marks the rise of a highly interconnected digital ecosystem. Often described as "computers on wheels," EVs operate within a complex network of EVC systems, cloud-based management platforms, and the power grid. While the integration of these hardware and software components enhances usability and promotes widespread adoption, it also creates a critical reliance on digital trust and cybersecurity. In this context, verifying *who* or *what* can access essential resources is no longer a mere technical function; authentication has become the cornerstone of digital trust in the advancing electromobility landscape (Chauhan et al., 2024).

The rapid emergence of new technologies has heightened the complexity of securing digital systems, placing authentication at the forefront of cybersecurity concerns. The EV ecosystem reflects broader trends seen in sectors like autonomous vehicles, the Internet of Things (IoT), and digital payments where trust is no longer grounded in physical presence but in a dynamic web of digital identities (Mo et al. 2022; Emodi et al. 2023). This review explores the evolution of authentication systems and underscores the urgent need for secure, scalable methods to authenticate vehicles, users, and globally distributed charging stations. To keep pace with evolving threats, the electromobility industry must move beyond static approaches like RFID and NFC, adopting intelligent, adaptive authentication frameworks capable of securing access across the interconnected network of EVs, EVCs, and the power grid.

Vulnerabilities in Traditional Authentication

Authentication has evolved significantly from physical tokens like keys to digital methods such as passwords. As digital ecosystems expanded, the limitations of static credentials became clear, giving rise to multi-factor authentication (MFA), which combines "something you know" with "something you have" or "something you are" (Lal, Prasad, and Farik 2016).

Authentication, fundamentally the process of verifying user identities and controlling access, is meant to safeguard systems while maintaining usability. Technologies like RFID and NFC emerged as key solutions to streamline interactions in smart, connected environments. Their early use in applications such as transit fare cards and building access systems demonstrated the power of convenience.

However, as smart devices proliferate, from industrial systems to EVs, the limitations of these static methods are increasingly exposed. For instance, many public EV charging stations rely on RFID-based authentication that transmits unencrypted, static identifiers. Researchers have demonstrated that these identifiers can be intercepted and cloned, allowing unauthorized access to charging infrastructure (Aubel et al., 2022). This highlights how traditional static mechanisms pose significant risks in the highly connected EV ecosystem. Another example is the weaknesses in the Open Charge Point Protocol (OCPP). Research has shown that some EV charging protocols like OCPP use static credentials for backend communication, increasing the risk of exploitation through credential theft or replay attacks (Hilt et al., 2021).

Designed primarily for speed and ease of use, RFID and NFC rely on transmitting unencrypted unique identifiers (UIDs) over short-range radio frequencies (Aubel et al., 2022). While standards such as ISO/IEC 14443 and NEMA EVSE 1-2018 enable interoperability, the fundamental communication model remains vulnerable to interception and spoofing. In today's digital landscape, effective authentication must go beyond convenience. There is a growing need for adaptive systems that dynamically respond to evolving threats, layer security based on real-time risk, and maintain a seamless user experience across platforms.

RFID and NFC Authentication Challenges in the EV Ecosystem

Without robust authentication mechanisms, the EV ecosystem remains exposed to a wide range of malicious activities including financial fraud, credential theft, and even physical sabotage of vehicles and charging stations (Hamdare et al. 2023). The continued reliance on static credentials such as RFID and NFC significantly increases the attack surface. These technologies carry inherent vulnerabilities that can be exploited to compromise vehicles, interrupt charging sessions, or inflict physical damage on infrastructure. As Joyce et al. (2023) explain, the core weakness lies in their broadcast nature; RFID, for instance, can unintentionally emit data well beyond its intended range, leaving it susceptible to passive eavesdropping even when the card remains inside a user's wallet. A compromised authentication system does more than inconvenience users; it has the potential to disrupt charging services, destabilize local power grids, and threaten the broader adoption and trust in electric mobility.

Without robust authentication mechanisms, the EV ecosystem is highly susceptible to a range of cyber and physical threats including credential theft, fraudulent energy consumption, and sabotage of charging infrastructure (Hamdare et al. 2023). Static authentication methods, such as RFID and NFC, are widely deployed across EV charging systems but present substantial security vulnerabilities. These systems typically rely on the unencrypted transmission of static UIDs, which can be intercepted using inexpensive tools. Once captured, these identifiers can be cloned or replayed, granting unauthorized access to charging stations or initiating fraudulent transactions.

As Joyce et al. (2023) emphasize, the core vulnerability of RFID lies in its broadcast nature; signals can propagate far beyond the intended range, making them vulnerable to passive eavesdropping even when a card remains in a user's wallet. Furthermore, because these systems often lack mutual authentication or dynamic cryptographic handshakes, attackers can impersonate legitimate devices or users without detection. These risks are not merely theoretical, researchers have demonstrated successful attacks that disrupt charging sessions, manipulate billing data, and even gain unauthorized control of EV infrastructure. A compromised authentication layer has cascading effects, potentially destabilizing load management systems, interrupting grid operations, and undermining public trust in the reliability of electric mobility.

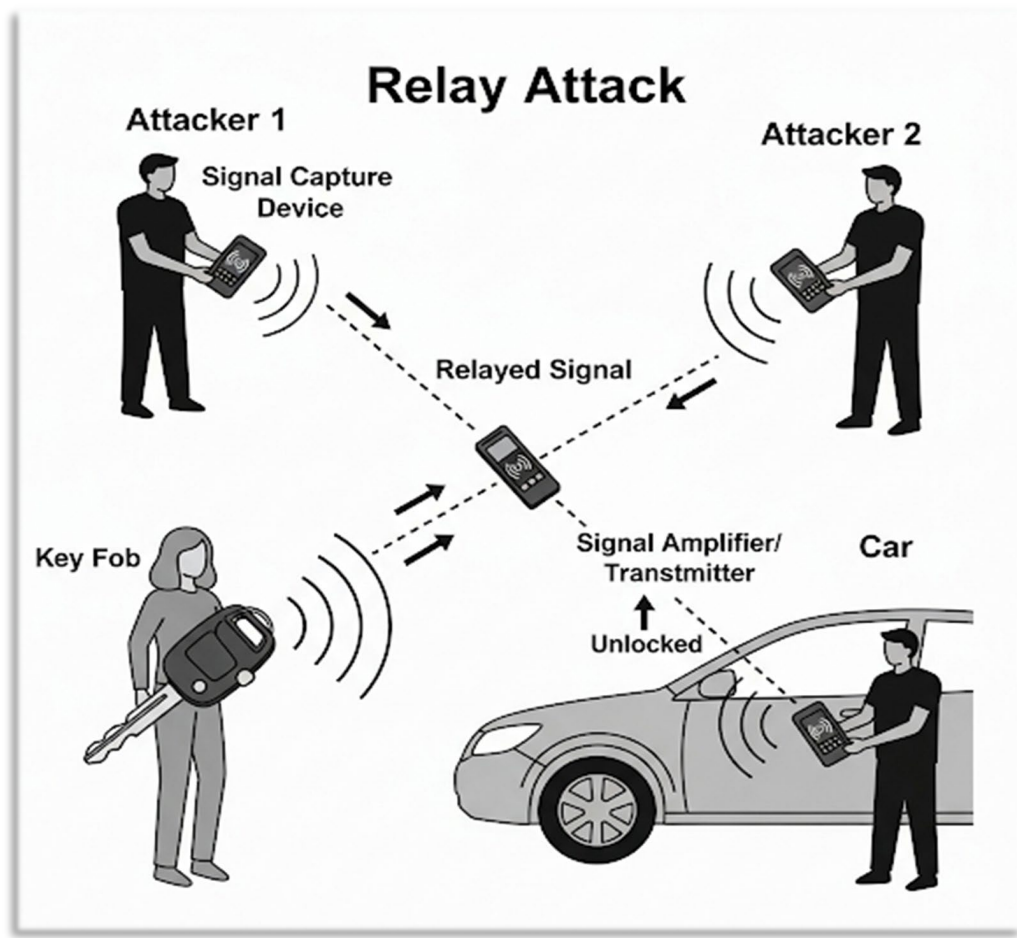

Figure 1. An example of RFID and NFC relay broadcast vulnerability (Gemini, 2025)

Limitations of RFID and NFC-Based Authentication in the EV Ecosystem

Authentication plays a critical role in the secure operation of EV charging infrastructure, particularly as vehicles, users, and grid systems become increasingly interconnected. Among the most widely implemented authentication methods in current EV charging systems are RFID and NFC technologies. These systems rely on radio waves to enable EVs to be charged wirelessly, without requiring physical contact between the vehicle and the charging station. In typical implementations, the vehicle identification and charging information are embedded within an RFID tag, which facilitates mutual recognition between the EV and the charging infrastructure (Joyce et al. 2023).

While these technologies offer convenience and speed, they also introduce a significant attack surface due to their reliance on static identifiers and unencrypted communications. RFID and NFC systems commonly transmit fixed UIDs that are broadcast via radio waves and can be intercepted by attackers using inexpensive tools such as software-defined radios or RFID sniffers. Once intercepted, these credentials can be cloned or replayed to gain unauthorized access to charging stations or initiate fraudulent transactions (Hamdare et al. 2023). The vulnerability is exacerbated by the passive broadcast nature of RFID. Joyce et al. (2023) note that RFID signals can unintentionally radiate far beyond their expected range, making them susceptible to passive eavesdropping even when cards remain inside a user's pocket or wallet. Furthermore, many current implementations of RFID-based authentication lack dynamic encryption, challenge-response protocols, or mutual authentication making them relatively easy to spoof. These security weaknesses are not just theoretical; researchers have demonstrated real-world attacks that exploit these flaws to disrupt charging sessions, impersonate EVs, or manipulate billing records (Aljohani and Almutairi 2024).

The consequences of a compromised authentication layer in the EV ecosystem are far-reaching. Beyond individual financial fraud or unauthorized charging, large-scale exploitation could destabilize load-balancing operations, interfere with grid-connected smart charging infrastructure, and erode consumer trust in electric mobility (Hamdare et al. 2023). As the EV industry scales globally, the limitations of static, radio-based authentication mechanisms highlight the urgent need for adaptive, cryptographically secure, and context-aware alternatives.

<div align="center">RFID and NFC Authentication Challenges</div>

RFID and NFC have become the dominant authentication mechanisms across the EV ecosystem, offering the promise of "contactless convenience" through a simple tap of a card or device. Automotive manufacturers including Tesla, BMW, and Hyundai have widely adopted these technologies due to their ease of use, speed, and reliance on physical proximity. However, this convenience masks critical security vulnerabilities. RFID and NFC protocols often transmit unencrypted identifiers, making them susceptible to a range of attack vectors such as cloning, eavesdropping, and relay attacks. These weaknesses create an expanded threat surface that adversaries can exploit to compromise vehicle access, disrupt charging sessions, or even stage coordinated physical intrusions (Zaidi et al. 2016).

<div align="center">Attack Vectors Exploiting Authentication in EV Ecosystem</div>

As electric vehicles (EVs) and their supporting infrastructure become increasingly integrated into intelligent transportation systems, secure authentication between EVs, charging stations (EVSE), and backend services has become a critical concern. These systems often rely on wireless protocols such as RFID, NFC, and Bluetooth, which, while enabling user convenience and automation, expose a broad attack surface. Authentication mechanisms in the EV ecosystem are particularly vulnerable to a range of cyber-physical threats that exploit the dynamic, distributed, and wireless nature of their communication. Adversaries can leverage both passive and active attack vectors such as signal interception, relay attacks, data manipulation, and credential spoofing to undermine authentication, gain unauthorized access, or disrupt services. Understanding these attack vectors is essential to designing resilient authentication frameworks that can withstand evolving threats in this high-mobility, cyber-physical environment (Hamdare et al. 2023).

*Signal Interception*

Signal interception is a foundational threat in wireless authentication systems, particularly in electric vehicle (EV) environments where RFID, NFC, and Bluetooth protocols are frequently used. As illustrated in Figure 2, this passive attack involves the unauthorized capture of communication signals transmitted between an EV, charging station, or backend server during authentication exchanges. Intercepted data such as identifiers, authentication tokens, or session keys can later be used to launch more sophisticated attacks, including relay, spoofing, or credential forgery. The broadcast nature of wireless protocols makes them inherently susceptible to interception, especially when encryption or mutual authentication is weak or absent. As EV adoption accelerates, the ability to silently capture and reuse authentication signals poses a growing threat to system integrity and user privacy (Joyce et al. 2023).

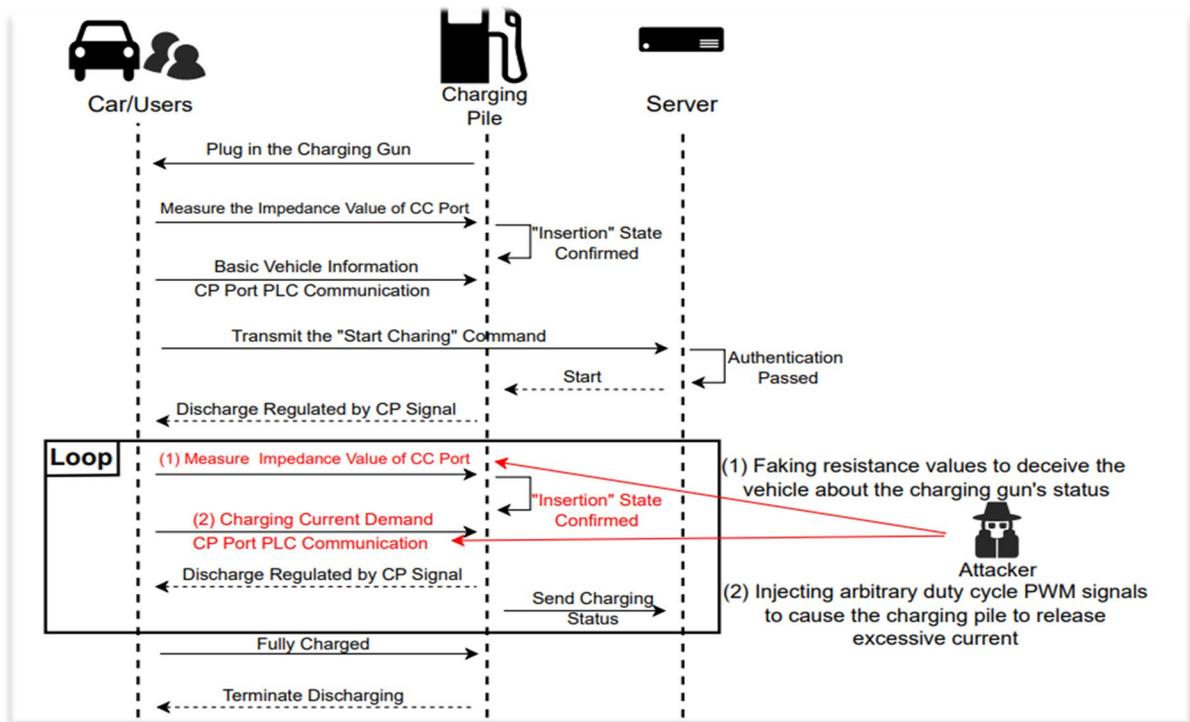

Figure. 2: Falsified Signal Attack Exploiting Weak Authentication in Charging Protocols (Shi et al. 2025)

*Eavesdropping*

Threat actors can exploit unencrypted communication between an RFID or NFC card and a charging station by using high-gain antennas to passively intercept data transmissions. This allows them to "listen in" on the credential exchange process, capturing unique identifiers or authentication tokens even from a distance without the user's knowledge (Hamdare et al. 2023).

*Relay Attacks*

These sophisticated attacks involve two colluding malicious devices: one positioned near the victim's authentication card and the other near the EV or charging station. The first device captures the card's signal and relays it in real-time to the second, effectively "tricking" the system into believing that the legitimate user is physically present. This technique allows attackers to unlock vehicles or initiate charging sessions remotely, without ever compromising the card itself (Almuhaideb and Algothami 2022).

*Data Manipulation and Signal Jamming*

Malicious actors can actively interfere with NFC transactions by altering data in transit modifying authentication parameters, redirecting payment information, or corrupting credential exchanges. Signal jamming, on the other hand, leverages electromagnetic interference to disrupt communication between the EV and the charging station, effectively launching a Denial-of-Service (DoS) attack that prevents users from authenticating or initiating charging sessions (Onumadu and Abroshan 2024).

*Replay and Spoofing*

In a replay attack, adversaries capture a legitimate authentication exchange and retransmit it at a later time to gain unauthorized access. Spoofing takes this further by imitating a valid RFID or NFC tag's signal to deceive the reader into accepting a fraudulent credential (Conti et al. 2022).

*Cloning and Credential Forgery*

Many RFID tags broadcast static Unique Identifiers (UIDs), making them highly susceptible to cloning. Attackers can extract the UID from a legitimate card and replicate it onto a blank tag, effectively creating a functional duplicate. This cloned credential grants unauthorized access to charging services and may even compromise user accounts. As Mitrokotsa et al. (2021) emphasize, "RFID tag skimming has enabled simple tools to become potent vectors for large-scale social engineering and denial-of-service campaigns against critical infrastructure" (Aubel et al. 2022; Mitrokotsa et al. 2021).

*Physical Layer Exploits*

Physical layer exploits target hardware interfaces and embedded components integral to authentication systems in electric vehicles and charging infrastructure. A compelling example, illustrated in Figure 3, is the *PHYSICAL-Layer Signal Injection Attacks on EV Charging Ports* study, where researchers demonstrated how attackers could insert a malicious device into the charger connector (branded as "PORTulator") to inject spoofed electrical signals. This approach enabled them to bypass authentication checks, induce charger lockouts, trigger denial-of-service conditions, and even manipulate in-vehicle CAN-bus messages across multiple charging standard, including SAE J1772, CCS, IEC 61851, and North American Charging Standard (NACS) (Shi et al. 2025). Such attacks bypass upper-layer cybersecurity and illustrate how vulnerabilities rooted in physical signaling mechanisms can completely undermine secure authentication.

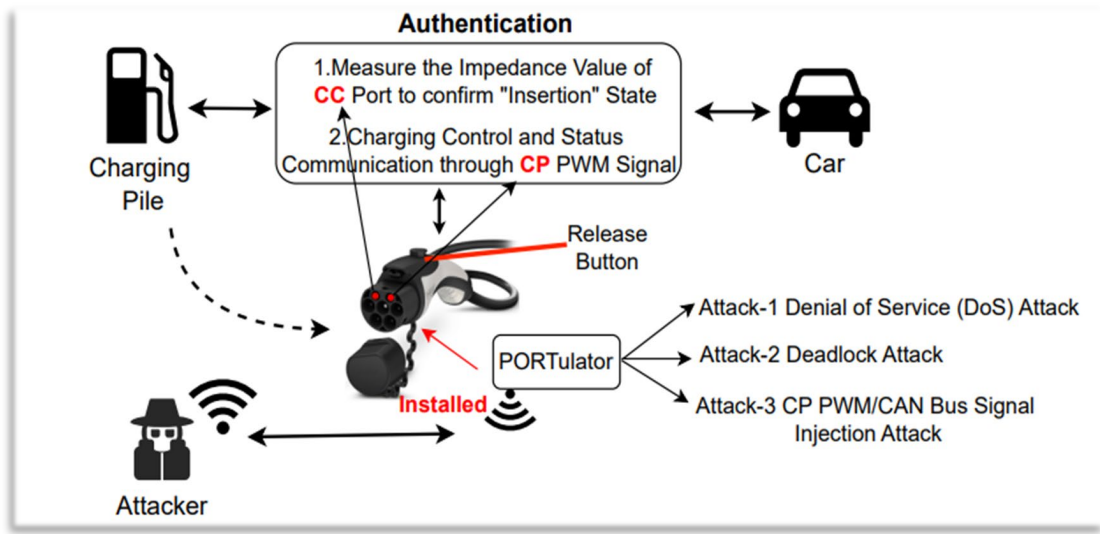

Figure 3: Overview of PORTulator Attack Vectors on EV Charging Infrastructure (Shi et al. 2025)

The impact of these vulnerabilities is summarized in the Table 1 below. A single authentication failure can trigger cascading effects across the interconnected EV ecosystem, compromising not only individual user access but also the integrity of charging infrastructure and, in some cases, the broader stability of the power grid.

Table 1: Common Vulnerabilities and Impacts in EVs and EVCs

| Vulnerability Category | Attack Mechanism | Impact |
|---|---|---|
| Signal Interception | Eavesdropping, Man-in-the-Middle (MiM) attack | Credential theft, unauthorized access, data manipulation |
| Relay-Based Attacks | Relay attacks forwarding, communication between card and reader | Remote unlocking, power theft, fraudulent charging |
| Data Manipulation & Disruption | Signal jamming, firmware-level exploits | Denial of Service (DoS), charging interruption, system crashes |
| Credential Forgery | RFID skimming, cloning of tags/cards | Unauthorized access, vehicle theft, account compromise |
| Physical Layer Exploits | Control Pilot spoofing, charging signal tampering | Charger lockout, overcharging, grid destabilization |

Toward AI-Powered Adaptive Authentication

Recognizing the inherent limitations of static authentication methods, adaptive authentication, also known as risk-based authentication, has emerged as a transformative solution. Unlike traditional one-size-fits-all approaches, adaptive systems continuously evaluate contextual signals and dynamically adjust security measures based on real-time risk assessments (Chauhan et al. 2024). This context-aware model enhances both security and user experience by tailoring authentication requirements to each unique access attempt. Key risk indicators such as login location, device fingerprinting, time of access, and behavioral patterns are analyzed to generate a risk score, which then determines the appropriate level of security intervention (Hitarth and Mahak 2025). When risk is minimal, access is granted seamlessly; when anomalies are detected, the system may invoke multi-factor authentication, escalate verification steps, or deny access altogether. This dynamic response capability strengthens system resilience and helps defend against evolving attack vectors.

AI-Powered Adaptive Authentication

As artificial intelligence (AI) continues to expand across critical infrastructure domains, it plays an increasingly vital role in enhancing cybersecurity frameworks. At the forefront of this evolution are AI-powered adaptive authentication systems, which move beyond the rigid, binary logic of traditional methods. Instead of relying on static credentials, these systems harness machine learning algorithms to analyze contextual signals, assess real-time risk, and adjust authentication protocols dynamically. By learning from large, continuously evolving datasets, they can detect anomalies, recognize behavioral patterns, and uncover hidden correlations, enabling a proactive, multi-layered defense posture that evolves alongside emerging threats (Chauhan et al. 2024).

Zero Trust Architecture as a Backbone for Adaptive Authentication

Zero Trust Architecture (ZTA) is built on key foundational principles that eliminate implicit trust and enforce strict security at every point of interaction. Rejecting the outdated concept of a "trusted internal network," the ZTA model operates on the principle of *"never trust, always verify"* (NIST 2020). For critical systems such as EVs and EV charging infrastructure, Zero Trust provides a framework that shifts security from a perimeter-based defense to a more granular, data-centric model. Its implementation is guided by several core security objectives:

- **Deny by Default:** All requests to network resources are blocked unless explicitly permitted by policy. Access is granted only when verified against strict user, device, or application rules.
- **Least Privilege Access:** Each identity is limited to the minimum set of permissions required to perform its function. This reduces the attack surface and mitigates the damage a compromised account can cause.
- **Authentication and Authorization:** Continuous, explicit verification is required for every access attempt. Authentication ensures the legitimacy of the entity, while authorization enforces policy-driven access where denial is the default.
- **Secure Communication:** All data in transit between endpoints, such as EV chargers and management systems, must be encrypted and integrity-checked to prevent interception or tampering.
- **Minimize Network Exposure:** Connectivity is restricted to essential functions. Techniques such as micro-segmentation create isolated zones that limit lateral movement by adversaries.

- **Trusted Network Infrastructure:** Critical services, including time synchronization (NTP) and DNS, must rely on operator-controlled, trusted infrastructure. This reduces dependency on untrusted third-party services that could otherwise be exploited (Carroll, Chang, and Wright-Hamor 2024).

    In summary, ZTA strengthens AI-powered adaptive authentication by providing the policy

backbone necessary for continuous verification in EV and EVC environments. Its principles, such as deny by default, least privilege, and secure communication, ensure that adaptive authentication mechanisms operate within a consistent, system-wide security framework. By aligning AI-driven, context-aware authentication with Zero Trust's granular and data-centric security controls, EV infrastructure can achieve dynamic resilience against evolving cyberattacks. This synergy minimizes network exposure, reduces reliance on vulnerable trust assumptions, and enables a more proactive, scalable defense for critical charging and mobility systems. Building on this foundation, the next section examines the specific mechanisms of AI-powered adaptive authentication, showing how real-time data analysis, machine learning, behavioral analytics, and context awareness operationalize Zero Trust principles in practice.

Mechanisms of AI-Powered Adaptive Authentication

As traditional authentication mechanisms struggle to keep pace with evolving threats, AI-powered adaptive authentication introduces a dynamic and intelligent approach to securing access in distributed and high-risk environments such as the electric vehicle (EV) ecosystem. As illustrated in Table 2, these systems leverage real-time data, machine learning algorithms, and contextual awareness to assess risk and make nuanced authentication decisions. By moving beyond static credentials, adaptive authentication integrates mechanisms such as context-aware risk scoring, behavioral biometrics, anomaly detection, continuous authentication, and automated threat mitigation. Each of these components contributes to a layered defense strategy that adapts to user behavior and environmental factors, reducing the attack surface while maintaining usability (Hamdare et al. 2023).

*Context-aware Risk Scoring*

This approach dynamically evaluates the risk level of an authentication attempt by analyzing contextual attributes such as the user's geographical location, login timestamp, originating IP address, device security posture, and historical behavioral patterns. By incorporating these factors, the system computes a composite risk score that can inform adaptive access decisions in real time (Hossen et al., 2025).

*Behavioral Biometrics*

Behavioral biometrics focuses on identifying users based on unique interaction patterns, such as mouse dynamics, keystroke rhythms, touchscreen gestures, or the sequence and timing of physical interactions at EV charging stations. Unlike traditional - static credentials, these behavioral traits are difficult to replicate, offering a passive and continuous form of user authentication (Sturgess et al. 2023).

*Anomaly Detection*

Anomaly detection leverages machine learning algorithms, ranging from traditional classifiers like the DecisionTreeClassifier to advanced neural networks, to model a baseline of legitimate authentication behavior. Once trained, the system detects outliers, such as an access attempt occurring immediately after multiple failed logins or from an unrecognized device. Such deviations trigger alerts or defensive actions, enabling real-time threat mitigation (Kuraku, 2023).

*Continuous Authentication and Automated Threat Migration*

This approach combines behavioral biometrics and predictive analytics to enable real-time, adaptive security throughout a user session. Rather than relying solely on a one-time login, the system continuously evaluates risk signals such as user behavior, device integrity, and environmental context to determine the trustworthiness of ongoing interactions. Based on the assessed risk level, proportionate responses are dynamically applied. For instance, low-risk sessions proceed with little or no friction; moderate-risk sessions trigger step-up authentication requiring multi-factor authentication; and high-risk sessions may be immediately blocked with automated security alerts issued. This layered, context-driven strategy ensures proactive threat detection and mitigation, particularly in environments like EV ecosystems where vehicles, charging stations, and grid infrastructure interact in complex, high-stakes ways (Kumari et al., 2024).

Table 2: Traditional vs Adaptive Authentication mechanisms in EV Ecosystems

| Feature | Traditional (Static) Authentication | Adaptive (Intelligent) Authentication |
|---------|-------------------------------------|----------------------------------------|
| Examples | RFID cards at charging stations<br>Static credentials in OCPP | Behavioral biometrics<br>AI-based risk scoring<br>Context-aware MFA |
| Credential Type | Static UID, hardcoded passwords | Dynamic tokens, ephemeral credentials |
| Data Encryption | Often unencrypted or weakly protected | Encrypted end-to-end communication |
| Vulnerability Type | Cloning, replay attacks, credential theft | Requires complex threat modeling to bypass |
| Security Response | Fixed and passive | Context-aware, responsive to anomalies (e.g., unusual access location) |
| Usability vs Security | Prioritizes speed and convenience | Balances convenience with real-time risk detection |
| Scalability & Future Readiness | Limited scalability in complex, distributed systems | Designed for distributed, global, and evolving networks |
| EV Relevance | Used in early EV charging infrastructure | Needed for secure EV-EVC-cloud-grid integration |

Literature Review Summary

The literature review provided a comprehensive examination of AI-powered adaptive authentication frameworks, beginning with foundational definitions and expanding into their pivotal role in securing the evolving EV ecosystem. It traced the shift from traditional, static authentication methods such as RFID and NFC, which prioritized user convenience but exposed significant vulnerabilities to more dynamic, context-aware solutions. The review emphasized how AI-driven adaptive authentication mitigates these risks by continuously assessing contextual and behavioral signals, enabling real-time adjustments to security protocols. This intelligent, flexible approach is essential for preserving the integrity, reliability, and trustworthiness of increasingly interconnected EV systems. As electric mobility scales and interactions between vehicles, charging infrastructure, and the power grid become more complex, AI-powered adaptive authentication frameworks emerge as a foundational pillar for sustaining secure and seamless operations.

## Chapter 4: Solution

### Introduction

The literature review has highlighted the critical security vulnerabilities inherent in RFID and NFC-based authentication methods that currently dominate the electric vehicle (EV) ecosystem. These static and convenience-oriented mechanisms are highly susceptible to a range of sophisticated attacks including relay attacks, spoofing, and physical layer exploits that pose serious risks such as financial loss, user compromise, and infrastructure disruption. In response to these evolving threats, there is a pressing need to transition from fragile, static protocols to a more intelligent and adaptive approach. This chapter introduces an AI-powered adaptive authentication framework specifically designed for the EV landscape. By leveraging machine learning algorithms, the proposed framework enhances resilience, scalability, and contextual awareness, offering robust protection across users, vehicles, charging stations, and the broader power grid infrastructure.

### Analysis of the proposed solution: AI-powered Adaptive Authentication

The proposed solution is a multi-layered, AI-powered adaptive authentication framework that fundamentally departs from the static logic of traditional systems. Instead of relying on a single, fixed credential, this framework performs a continuous, real-time risk assessments for each authentication attempt and subsequent user interaction. At its core, the framework leverages machine learning to analyze a continuous stream of contextual and behavioral data, generating a dynamic risk score that dictates the appropriate level of security. This enables proportional security responses–offering a seamless, low-friction experience in low-risk situations, while escalating defenses in the presence of anomalies or suspicious behavior.

### AI-powered Adaptive Authentication as a Multi-layered Framework

The AI-powered adaptive authentication framework is fundamentally driven by layered machine learning (ML) techniques, which serve as the engine behind its dynamic and context-aware capabilities. Unlike traditional authentication systems that validate identity only at a single point of entry, this adaptive approach emphasizes continuous verification throughout the entire communication process. Leveraging the transformative power of AI, the system can process vast and heterogeneous datasets, detect subtle behavioral anomalies, and automatically trigger tailored security responses based on real-time contextual risk assessments. At the heart of this framework, ML enables granular access control, dynamic policy enforcement, and persistent validation of users, devices, applications, and transactions. Its strength lies in continuous learning monitoring live data streams to refine behavioral baselines and detect deviations that static, rule-based systems often overlook. For example, ML algorithms analyze patterns ranging from touchscreen interaction to a vehicle's electrical charging signature to establish personalized behavioral profiles. These algorithms perform real-time risk scoring by integrating multiple contextual signals and assigning a dynamic risk level to each authentication event. This transforms the security posture from reactive to proactive, allowing the system to anticipate and mitigate emerging threats before they materialize.

Furthermore, the integration of feedback loops where authentication outcomes are continuously fed back into the learning model enhances detection accuracy over time and reduces false positives. This iterative process ensures that the framework evolves in tandem with user behavior and threat landscapes, offering a more intelligent, responsive, and resilient security mechanism for the electric vehicle ecosystem (Jorquera et al. 2020; Hitarth & Mahak 2025).

Integrating Zero Trust Principles into an Adaptive Authentication Framework

The framework is fundamentally grounded in the principles of Zero Trust Architecture (ZTA) and continuous verification, representing a significant departure from traditional, static security models. Adopting ZTA is not merely an incremental improvement but a fundamental paradigm shift. Conventional security models, often perimeter-based, implicitly trust entities once they gain initial access to a network (Hitarth & Mahak 2025). This approach, exemplified by Role-Based Access Control (RBAC) and Attribute-Based Access Control (ABAC), relies on predefined rules or static attributes. While effective in structured environments, such reliance makes these models susceptible to modern identity-based attacks such as credential theft and lateral movement, allowing malicious actors to operate undetected once initial access is compromised. In contrast, the Zero Trust model operates under the principle of "never trust, always verify", ensuring that access is continuously assessed through contextual risk, user behavior, and real-time analytics. This philosophy eliminates implicit trust, enforcing continuous authentication, conditional access, dynamic trust evaluation, and the principle of least privilege (Kandula et al. 2024). This shift from a reactive, perimeter-based defense to a proactive, data-centric, and identity-aware posture is critical for securing complex, interconnected ecosystems like EV charging infrastructure. Within this model, adaptive authentication acts as a critical enabler for ZTA, dynamically adjusting the required level of authentication based on contextual risk. By continuously evaluating access through real-time analytics, ZTA ensures that trust is never static but constantly reassessed.

<div align="center">Machine Learning Model Architectural Framework</div>

The framework is fundamentally built upon layered machine learning techniques, which serve as the engine behind its adaptive capabilities. Its core strength lies in the ability to synthesize diverse data streams into holistic, context-aware intelligence, moving beyond isolated data points to form a comprehensive understanding of each access attempt (Ahmed et al. 2025). The architecture is composed of the following functional layers:

- **Data Collection**: This foundational layer aggregates data from diverse sources, including user activity patterns (e.g., charging times and locations), device trustworthiness (e.g., OS patch levels), network conditions, and historical usage behaviors. These inputs enable the system to establish dynamic, high-fidelity behavioral baselines.

- **Risk Assessment and Scoring**: The framework assigns a continuous risk score using a weighted combination of contextual signals. This converts subjective security judgments into objective, quantifiable metrics. Based on this computed risk score, the system selects an appropriate response aligned with predefined risk tiers—low, medium, or high (Ahmed et al. 2025).

- **Behavioral Analytics:** Machine learning algorithms continuously monitor and learn user behavior over time, creating detailed activity patterns. Attributes such as device fingerprinting, login frequency, IP geolocation, network reputation, and behavioral biometrics are monitored to enhance user-specific modeling (Jorquera et al. 2020).

- **Anomaly Detection:** This layer employs a combination of supervised (e.g., Random Forest, Adaptive Random Forest, Support Vector Machine), unsupervised (e.g., Isolation Forest, Autoencoder), and deep learning techniques (e.g., Gated Recurrent Units, Recurrent Neural Networks, Convolutional Neural Networks) to identify deviations from established behavioral baselines (Jorquera et al. 2020).

- **Automated Policy Enforcement and Threat Mitigation**: This final layer translates the computed risk score into concrete security actions—such as Allow, Challenge, or Deny—based on administrator-defined rules and policies. These dynamic actions ensure proportional response to varying threat levels and reduce user friction during legitimate interactions.

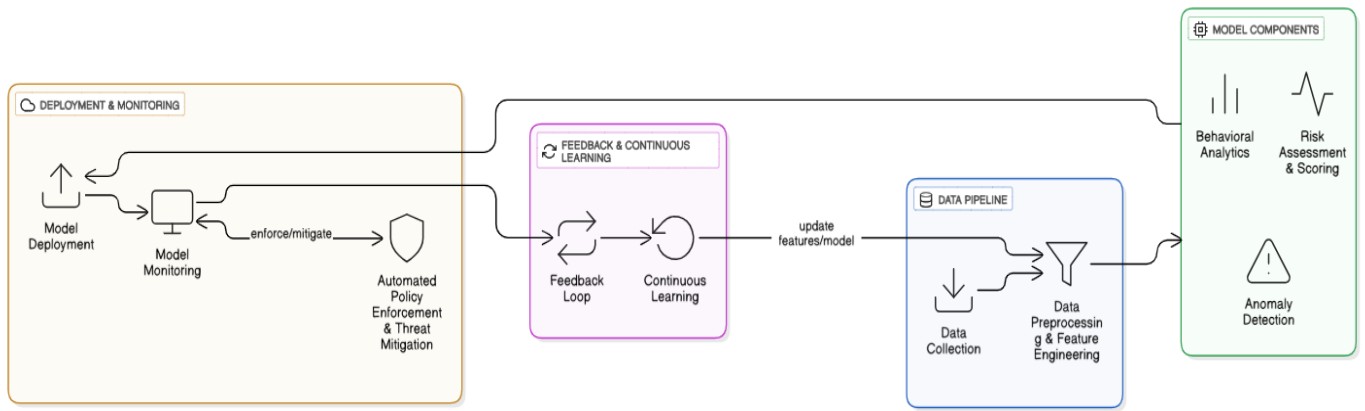

Figure 4. Machine Learning Model Architectural Framework (Gemini,2025)

Application of AI-powered Adaptive Authentication Framework in the EV ecosystem

The AI-powered adaptive authentication framework directly addresses the inherent vulnerabilities of traditional RFID and NFC-based systems by employing a multi-layered, intelligent approach that defends against a wide spectrum of attack vectors, surpassing the limitations of simple encryption. In the complex EV ecosystem, where users, vehicles, charging stations, and the power grid continuously interact, static authentication methods quickly become liabilities once credentials are compromised. These legacy systems lack the capacity to adapt to evolving contexts or recognize anomalous behavior (Anvesh 2022). For instance, if valid credentials are stolen, a static system will still grant access allowing attackers to move freely and undetected within a system. In contrast, the AI-powered framework continuously assesses contextual and behavioral signals such as device recognition, login location, and usage patterns. This dynamic risk assessment enables the system to detect anomalies like an unfamiliar device or unusual behavior and respond in real time by escalating authentication requirements or denying access altogether (CSA 2025). Even if credentials are compromised, unauthorized access can be intercepted and neutralized. The system adapts security requirements proportionally, ensuring a seamless, frictionless experience for low-risk scenarios while reserving challenges for elevated-risk situations. For EV charging stations, this adaptive intelligence improves detection accuracy, automates threat response, and reduces manual security workload ultimately enhancing both user experience and system resilience.

      In summary, the proposed solution is a multi-layered, AI-powered adaptive authentication framework that moves beyond the static logic of traditional systems. By integrating Zero Trust principles with adaptive authentication and machine learning model-driven analytics, the proposed framework provides continuous, context-aware, and resilient identity verification tailored to complex, dynamic environments such as EV charging infrastructures.

**Chapter 5: Discussion**

Introduction

Building on the insights from the literature review, this chapter examines the proposed solution: a multi-layered, AI-powered adaptive authentication framework tailored for electric vehicles (EVs) and electric vehicle charging (EVC) infrastructures. The comprehensive review highlighted that while traditional, static methods such as Radio-Frequency Identification (RFID) and Near Field Communication (NFC) offer users convenience, their static nature creates exploitable vulnerabilities. Responding to this problem, the proposed framework advances beyond single-point credential checks by integrating continuous risk evaluation, contextual analysis, and behavioral intelligence throughout the authentication lifecycle. A foundational component of this framework, as discussed in the paragraph below, is the use of machine learning (ML), which enables the system to detect anomalies, predict emerging threats, and adapt authentication requirements in real-time. Collectively, these capabilities are organized into four interdependent layers, the multi-layered architecture: continuous authentication, dynamic risk assessment, behavioral anomaly detection, and predictive threat detection that directly address the gaps and limitations identified in prior studies and strengthens digital trust within the expanding EV ecosystem.

The Intelligence Layer: Machine Learning in Adaptive Authentication

Machine Learning (ML) is a foundational component of adaptive authentication, serving as the core engine that enables systems to learn from user behavior, detect anomalies in real-time, dynamically adjust security measures. Its strength lies in its ability to continuously learn and evolve, allowing the system to establish baselines of typical user behavior from historical data and adapt to changing patterns over time. This adaptive capability is critical for distinguishing between legitimate behavioral shifts and potentially malicious activity, thereby reducing unnecessary friction for genuine users. ML excels at pattern recognition and anomaly detection, making it highly effective in identifying fraudulent attempts as they occur. It plays a central role in behavioral biometrics, analyzing unique interaction patterns such as keystroke dynamics, touchscreen gestures, or how a device is held to construct individualized behavioral profiles. Deviations from these profiles, such as unusual login times or attempts to access sensitive data, can signal possible compromise and trigger additional security measures.

ML algorithms are instrumental in enabling adaptive behavior by learning from historical data to establish baselines of typical user actions, which turn allows the system to dynamically adjust its security responses. This learning capability is crucial for distinguishing between genuine anomalies and benign shifts in user behavior, thereby minimizing unnecessary fiction for legitimate users.

ML is pivotal for behavioral biometrics and anomaly detection. It analyzes user behavior patterns, such as touchscreen interactions, and even how a user holds a device. By constructing unique behavioral profiles, ML models can discern deviations that might indicate a compromised account or the presence of a malicious actor. This includes flagging irregular access patterns, unusual login times or attempts to access sensitive data outside typical interactions.

Furthermore, ML algorithms serve as the primary mechanism for real-time risk scoring. By processing multiple contextual signals such as location, device fingerprinting, access patterns, and user history, they generate dynamic risk profiles for each login or transaction. These risk scores guide the system in tailoring authentication requirements, allowing seamless access in low-risk scenarios while escalating verification in high-risk ones.

Machine Learning Enabled Framework for Adaptive Authentication

The proposed solution is a multi-layered, AI-powered adaptive authentication framework that transcends the limitations of traditional static systems. Instead of relying on a single, static credential check at login, this framework continuously evaluates risk in real time, both at the initial point of authentication and throughout the user's session. At its core, the system ingests and analyzes a stream of contextual and behavioral data to calculate a dynamic risk score, which determines the appropriate level of authentication required at any given moment. This ensures that security measures are intelligently scaled to the threat level such as providing a seamless, low-friction experience for trusted users in low-risk contexts, while rapidly escalating defenses when suspicious patterns or anomalies are detected. To operationalize this proposed solution, the framework is structured into several interdependent layers, each addressing a distinct dimension of adaptive security.

The following subsections elaborate on its key components: continuous authentication and predictive analysis capabilities, risk assessment, behavioral anomaly detection, and predictive threat detection.

*Continuous Authentication and Predictive Analysis Capabilities*

A cornerstone of the framework is continuous authentication, which replaces one-time credential checks with ongoing verification process. By leveraging predictive analysis, the system adapts authentication requirements in real time, making sure that trust is dynamically maintained throughout a user's session.

The framework employs continuous monitoring and predictive analysis to safeguard user sessions and anticipate emerging threats. This approach represents a fundamental shift from traditional reactive security models to proactive threat mitigation. By persistently analyzing behavioral patterns, device integrity, geographic location, and network conditions, the system maintains real-time awareness of user context. When inconsistencies or anomalies are detected such as unexpected location changes or suspicious device behavior, the framework dynamically escalates security protocols to contain potential threats. This ongoing validation is critical for preventing lateral movement by malicious actors and minimizing the risk of insider threats, ensuring that access is secure throughout the session (Jimmy 2024).

The framework employs continuous monitoring and predictive analysis to maintain security throughout user sessions and anticipate emerging threats. This integration fundamentally shifts the security posture from reactive incident response to proactive threat mitigation. The continuous session monitoring ensures surveillance checks for consistency in user behavior, device status, geographical location, and network conditions. The system dynamically imposes stronger security measures if deviations are detected. This continuous validation is essential for mitigating risks such as lateral movement by malicious actors and reducing insider threats (Jimmy 2024).

This constant verification, continuous monitoring and predictive analysis, provides the foundation for dynamic risk assessment, where contextual and behavioral signals are analyzed to determine the appropriate level of trust.

*Risk assessment*

At the core of adaptive authentication lies a dynamic risk assessment engine that evaluates contextual, behavioral, and environmental signals. Based on the threat level, this process calculates and then generates a real-time risk score that guides the system in scaling authentication demands.

This component serves as the analytical heart of the framework, tasked with processing contextual and behavioral data to compute a real-time risk score for each authentication attempt. It employs advanced mathematical modeling and machine learning techniques to dynamically assess threat levels. As demonstrated by Hitarth and Mahak (2025), the dynamic risk score(R) is continuously computed using the following model

$R = αU + βD + γT + δf(t)$

Where:

- U represents a composite metric for user behavior (e.g., login time anomaliesgeolocation deviations).
- D reflects device integrity (e.g., operating system health, patchcompliance).
- T incorporates external threat intelligence scores.
- f(t) captures time-dependent risk fluctuations.
- α, β, γ, and δ are tunable weighting parameters optimized via cross-validation techniques.

This real-time risk computation enables the system to adaptively adjust authentication requirements based on the evolving threat landscape, ensuring that security measures remain both responsive and proportionate. However, risk scores alone are not sufficient. Detecting subtle shifts in user behavior is equally important, which is where behavioral anomaly detection becomes a critical next layer in the multi-layered architecture.

*Behavioral Anomaly Detection*

To strengthen proactive defense, the proposed framework incorporates behavioral anomaly detection that learns from user patterns over time. The system, by identifying deviations from normal behavior, can flag suspicious activity and trigger escalated authentication measures before damage occurs.

This key capability leverages machine learning techniques to detect abnormal user behavior during device interactions. The framework continuously compares current usage patterns against baseline profiles that are dynamically updated in real time. By measuring the degree of similarity between observed and expected behavior, the system can identify potential anomalies with high accuracy. Algorithms like Isolation Forest are particularly effective for this purpose, providing a quantifiable anomaly score that reflects the likelihood of malicious activity (Jorquera et al. 2020). While anomaly detection identifies deviations as they occur, the proposed framework further advances security posture through predictive threat detection, anticipating future risks before they materialize.

*Predictive Threat Intelligence*

Extending beyond immediate anomaly detection, the proposed framework employs predictive threat detection powered by machine learning models. These models predict potential attack vectors by analyzing both historical trends and emerging threat indicators, enabling the system to intelligently adjust defenses.

By analyzing vast volumes of both structured and unstructured data, AI enables the identification of patterns and the detection of anomalies that may signal an attack, even when the threat has no known signature. This capability marks a shift from reactive cybersecurity to a proactive defense posture, empowering systems to anticipate and neutralize threats before they can compromise networks or assets. AI can analyze network traffic, user behavior, and even linguistic patterns to uncover subtle indicators of emerging attacks. This foresight supports automated, preemptive actions such as blocking suspicious IP addresses, flagging abnormal activity, or isolating compromised devices, thereby reducing response time and limiting potential damage (Jimmy 2024).

Collectively, these interdependent layers, continuous authentication, risk assessment, behavioral anomaly detection, and predictive threat detection, form a cohesive, multi-layered architecture that delivers resilient, adaptive security protection for complex ecosystems such as EVs and EV charging infrastructures.

The Zero Trust Layer: Continuous Verification in Adaptive Authentication

The Zero Trust layer lies at the core of this adaptive authentication multi-layered framework which enforces continuous verification as its foundational principle. This ensures that trust is never assumed but always earned in real-time during every interaction within the electric vehicle ecosystem.

In this framework, ZTA acts as the policy-driven core while AI serves as the enabler for real-time, context-aware decision-making. The synergy between ZTA and adaptive authentication is most evident in its ability to continuously verify every session and activity. Instead of trusting a user or device after a single point of authentication, this integrated model validates every access request and treats each interaction as if it originates from an untrusted network or source.

The principles of ZTA form a rigorous framework for continuous verification in the EV ecosystem. The *deny by default* principle blocks all access unless explicitly permitted by policy. This complemented by the principle of *least privilege* which limits each identity's permissions to only what is essential for its role, thereby reducing potential attack from a compromised account. *Authentication and authorization* require ongoing validation with policy enforcement while *secure communication* mandates the encryption and integrity verification of all data in transit while minimizing interception risks in EV charging systems (Oluoha et al.,2024).

Zero Trust Integration within the Adaptive Authentication Framework

A defining component of the proposed framework is its integration of Zero Trust Architecture (ZTA) with AI-powered adaptive authentication. This fusion redefines security by embedding ZTA's rigorous verification principles into an AI-powered system that continuously assesses and adapts to risks. This creates a proactive defense mechanism tailored for interconnected systems like EV charging networks. The convergence of Zero Trust Architecture and AI-powered adaptive authentication creates a security model that is greater than the sum of its parts. This integration forms a symbiotic relationship where ZTA provides the strategy for access control, and adaptive authentication acts as the dynamic enforcement mechanism that brings that strategy to life (Carroll et al., 2024).

ZTA serves as the policy framework that defines the security attribute of an adaptive authentication system. It is within the ZTA Policy Engine that a system solidifies the "why" and "what" of its access control strategy. These policies are based on risk tolerance, data classification schemes, and compliance requirements. These policies are enforced through the core principles which play out in different scenarios. The principle of *Deny by Default* ensures that all requests to network resources including the mobile application, an EV charger connecting to a charger, or a charger communicating with the backend are blocked unless explicitly permitted by a policy. The principle of *Least of Privilege* access limits each identity to the minimum set of permissions. For instance, a user's account should only authorize and monitor their charging session without having access to administrative functions. The principle of secure communication mandates that all data in transit between endpoints including the EV, the charger and the application management system must be encrypted to prevent eavesdropping during charging session (Kandula et al.,2024).

Zero Trust Architecture shifts away from perimeter-based models by assuming no implicit trust and requiring validation of every access request following these principles. As illustrated in the Figure 5 diagram below, when Zero Trust principles are integrated with AI-powered adaptive authentication, AI enhances these tenets by analyzing real-time signals, including behavioral patterns, environmental attributes, and threat intelligence to modulate authentication strength. For instance, if a user attempts to authenticate with an RFID card, the AI-powered adaptive authentication system assesses contextual factors like consistency of charger location with the vehicle's GPS data, charging times, and charging card usage. Based the calculated risk score and policies, the system can make a decision aligning with ZTA's "never trust, always verify" mantra (Carroll et al., 2024).

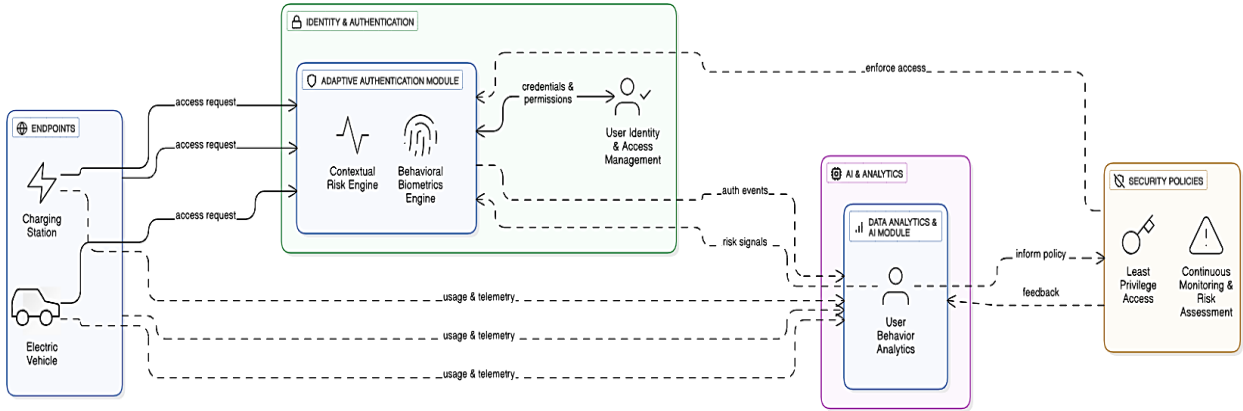

Figure 5. Overview of AI-powered Adaptive Authentication Framework with Zero-Trust Architecture (Gemini, 2025)

Securing the Electric Vehicle Ecosystem with AI-powered Adaptive Authentication

Within the EV ecosystem, adaptive authentication extends beyond generic data input by leveraging EV-specific datasets to enhance contextual risk assessment. ML models are trained on parameters such as charging frequency, duration, energy consumption patterns, preferred charge levels, and typical charging times, data that reflects user-specific charging behavior. In addition, these models can incorporate detailed driving characteristics, including acceleration patterns, speed profiles, driving style (e.g., conservative, moderate, or aggressive), and frequently traveled routes. Interactions with charging infrastructure such as historical access to specific charging stations and associated network characteristics further enrich the behavioral profile. By analyzing these multidimensional datasets, the system constructs a dynamic and highly contextualized baseline of normal behavior for both the user and the vehicle, enabling more precise anomaly detection and real-time threat mitigation.

In the EV ecosystem, adaptive authentication moves beyond generic data points by training on EV-specific datasets such as charging times, frequency, duration, energy consumption, and preferred charge levels. Beyond charging, ML models can integrate detailed driving behaviors, including driving style (conservative, moderate, aggressive), speed profiles, acceleration patterns, and common routes. Furthermore, interactions with charging infrastructure, such as historical access patterns to specific charging stations, network characteristics, provide invaluable data for comprehensive risk assessment. By analyzing these datasets, ML models can establish a highly contextualized baseline of normal behavior for both the user and the vehicle (Kumari et al. 2024).

In summary, the discussion has discovered that a multi-layered, AI-powered adaptive authentication framework, grounded in Zero Trust principles, provides a comprehensive response to the vulnerabilities inherent in traditional static security models. By integrating continuous authentication, dynamic risk assessment, behavioral anomaly detection, predictive threat detection, and machine learning intelligence, the proposed framework provides a scalable, context-aware approach to strengthening the security of electric vehicles (EVs) and electric vehicle charging (EVC) infrastructures. The integration of Zero Trust principles and AI-driven adaptive authentication represents a strong paradigm shift in security digital mobility, moving from static, reactive defenses to intelligent, predictive, and continuously verified trust in real-time. Building on these insights, the next chapter presents recommendations for implementation, policy alignment, and future research necessary to further operationalizing this proposed framework within the evolving electric mobility landscape.

## Chapter 6: Recommendations

The evolving cybersecurity landscape facing electric vehicles (EVs) and electric vehicle charging systems (EVCs) requires both innovation and discipline in the way new security frameworks are deployed. As this study has shown, current traditional authentication mechanisms such as Radio-Frequency Identification (RFID) and Near-Field Communication (NFC) provide only limited security protection and therefore remain vulnerable to cloning, spoofing, and man-in-the-middle attacks. In contrast, AI-powered adaptive authentication offers a transformative pathway toward more secure, scalable, and intelligent protection. Yet, its successful adoption depends on bridging theoretical research with real-world deployment. To achieve this, careful consideration must be given to phased implementation, robust data infrastructure, end-to-end security design, and regulatory alignment.

Recommendations for Deployment and Implementation

The findings of this qualitative study revealed opportunities for future research and practical innovation in the field of EV authentication. Drawing from the analysis of numerous peer-reviewed literature sources, the study highlighted the persistent vulnerabilities of traditional RFID and NFC-based authentication mechanisms and the transformative potential of AI-powered adaptive authentication systems. This research contributes to a growing body of knowledge that encourages multidisciplinary exploration of cybersecurity, machine learning, and behavioral biometrics in the EV ecosystem. Based on the identified gaps, limitations, and emergent trends, the following recommendations provide a roadmap for academies, automakers, infrastructure providers, cybersecurity professionals, and policymakers to translate research findings into actionable strategies for safeguarding the EV ecosystem:

*Practical Deployment through Pilot Programs*

The deployment of AI-powered adaptive authentication should begin with controlled pilot programs. Pilot testing in limited environments provides the opportunity to validate technical feasibility while reducing risk exposure. These pilots provide an opportunity for developers to measure system performance against real-world conditions, refine decision thresholds, and identify unforeseen challenges that may not surface during simulation or lab-based testing. Pilot deployments also generate valuable operational feedback, enabling iterative improvements in authentication models. By starting small, organizations can build confidence among stakeholders, address practical barriers such as latency or usability, and establish clear benchmarks for scalability. Ultimately, pilot programs serve as a critical proving ground, balancing innovation with pragmatism and ensuring that AI-power adaptive authentication frameworks can function reliably before widespread adoption.

*Data Collection Infrastructure and Readiness*

The effectiveness of adaptive authentication depends heavily on access to high quality, real-time data. Hence, robust data pipelines must be established to ensure consistent and secure ingestion of key data inputs, including user behavioral patterns, charging frequency, location history, device specifications, and network conditions. EV stakeholders must invest in infrastructure that enables secure and consistent ingestion of diverse data sources. Relevant inputs may include user behavioral patterns, charging frequency, geolocation data, device specifications, and network conditions. This data must be processed in real-time and supported by preprocessing mechanisms that ensure data quality, completeness, and relevance. Additionally, privacy-preserving technologies such as differential privacy and secure multiparty computation should be considered to balance user confidentiality with the need for high-fidelity data. Building on this foundational layer is critical not only for training and inference but also for enabling the continuous learning loops that allow adaptive authentication systems to evolve in response to emerging threats.

*End-to-End Security and Protocol Hardening*

AI-powered adaptive authentication cannot operate in isolation; it must be embedded within a secure end-to-end system architecture. Security-by-design principles should guide every stage of deployment, ensuring that authentication, telemetry, and model reference are safeguarded against compromise. All communications should be secured with Transport Layer Security (TLS) protocols with strong cipher suites and enforced certificate validation. Where feasible, TLS client certificates should be deployed to strengthen mutual authentication between EVs, charging stations, and management systems. Additional hardening measures, such as micro-segmentation and intrusion detection systems, can further reduce the risk of lateral movement across networks. Embedding these safeguards from the outset ensures that adaptive authentication does not become a single point of failure but rather a reinforcing layer within a resilient security ecosystem.

*Policy and Standards Alignment*

The sustained success of AI-powered adaptive authentication in EV and EVC systems depends on harmonizing technical innovation with cohesive standards and governance frameworks. As Morisset et al. (2021) underscore, aligning with common security standards and fostering cross-sector collaboration among industry, governmental, and international stakeholders is vital for robust EV charging cybersecurity. Automakers, infrastructure providers, and policymakers should jointly engage with standards bodies to embed adaptive authentication principles, risk-based access control, and Zero Trust approaches into EV authentication protocols. At the same time, regulatory bodies can stimulate adoption through cybersecurity mandates and incentives that encourage interoperability and secure design practices. This dual strategy would fortify resilience across the ecosystem and ensure that solutions gain scalability and global coherence.

In conclusion, these recommendations highlight the multi-dimensional approach required to deploy AI-powered adaptive authentication effectively within EV and EVC ecosystems. Pilot programs enable phased adoption and provide real-world validation of technical feasibility. Robust data infrastructure ensures that machine learning models have the quality and volume of inputs necessary to perform effectively. Security-by-design principles, reinforced through end-to-end encryption and protocol hardening, guarantee that adaptive authentication strengthens rather than weakens overall system resilience. Finally, policy alignment and standards development provide the governance structures required for long-term sustainability, interoperability, and trust. Together, these proposed recommendations bridge the gap between academic research and operational practice, equipping stakeholders with a roadmap to safeguard EVs and charging systems against increasingly sophisticated cyber threats. By integrating AI-powered adaptive authentication within a Zero Trust framework, the EV ecosystem can embrace innovation without compromising on the core principles of security, privacy, and reliability.

## Chapter 7: Conclusion

The rapid adoption of Electric Vehicles (EVs) and the corresponding growth of Electric Vehicle Charging Systems (EVCs) have created unprecedented opportunities for sustainable mobility but also introduced complex cybersecurity challenges. Authentication remains one of the most critical aspects of protecting vehicles, users, and the broader energy infrastructure. While technologies such as RFID and NFC have provided convenient, standardized methods for vehicle access and charging authorization, their reliance on weak cryptographic protocols and susceptibility to attacks including cloning, relay attacks, spoofing, and eavesdropping render them inadequate in the dynamic threat landscape of the EV ecosystem. These vulnerabilities highlight the urgent need for more resilient, adaptive, and intelligent authentication strategies.

This study has argued that AI-powered adaptive authentication offers a viable and scalable solution to these challenges by integrating machine learning, anomaly detection, behavioral analytics, and contextual awareness into the authentication process. Unlike traditional static methods, adaptive authentication responds dynamically to evolving risks, enabling proactive defense against cyberattacks. When implemented within the broader framework of Zero Trust Architecture, these mechanisms ensure that every access request is continuously verified, permissions remain strictly limited, and communication across EV ecosystems is safeguarded. This multi-layered approach provides a forward-looking model for strengthening resilience across EVs and EVCs while maintaining usability and efficiency for end users.

Ultimately, the proposed AI-powered adaptive authentication framework contributes to the development of secure digital identities within the electric mobility ecosystem. It reinforces the principle that as EVs continue to evolve into "computers on wheels," their security must evolve in parallel, guided by intelligent, adaptive, and Zero Trust–aligned solutions. Beyond advancing technical knowledge, this research underscores the broader imperative of aligning cybersecurity with innovation, ensuring that the benefits of electrification and smart mobility are realized without compromising trust, privacy, or critical infrastructure stability. Embracing AI-driven adaptive authentication is not merely a technical upgrade; it is a strategic imperative for securing the future of electric mobility.